\documentclass[]{jkas} 

\def\year{2023} 
\def\volume{56} 
\def\issue{1} 
\def\beginpage{1} 
\def\received{April 28, 2026} 
\def\accepted{July 05, 2026} 
\def\published{July 20, 2026} 
\date{Received \received; Accepted \accepted; Published \published}

\newcommand\ion[2]{{#1}\,\textsc{#2}} 

\title{%
A Search for Exoplanets around Northern Circumpolar Stars X.\\
The origin of radial velocity variations in the evolved star HD~216595
}

\author[1]{Sang-Hee Kim}{}
\author[2,3 $\star$]{Byeong-Cheol Lee}{0000-0002-2783-0755}
\author[4,5]{Shenghong Gu}{0009-0004-3025-3444}
\author[2]{Jae-Rim Koo}{0000-0001-8969-0009}
\author[6]{Beomdu Lim}{0000-0001-5797-9828}
\author[1]{Myeong-Gu Park}{0000-0003-1544-8556}
\author[7]{Huan-Yu Teng}{0000-0003-3860-6297}
\author[2]{Yeon-Ho Choi}{0009-0005-5145-5165}
\author[8]{David Mkrtichian}{0000-0001-5094-3910}
\author[1]{Tae-Yang Bang}{0009-0008-2592-9437}
\author[2]{Hyeong-Ill Oh}{}
\author[1]{Heon-Young Chang}{0000-0003-2015-2725}

\affil[1]{Department of Astronomy and Atmospheric Sciences, Kyungpook National University, Daegu 41566, Korea}
\affil[2]{Korea Astronomy and Space Science Institute, 776, Daedeokdae-Ro, Youseong-Gu, Daejeon 34055, Korea}
\affil[3]{Korea University of Science and Technology, Gajeong-ro Yuseong-gu, Daejeon 34113, Korea}
\affil[4]{School of Astronomy and Space Sciences, University of Chinese Academy of Sciences, Beijing 100049, People’s Republic of China}
\affil[5]{Yunnan Observatories, Chinese Academy of Sciences, Kunming 650216, People’s Republic of China}
\affil[6]{Department of Earth Science Education, Kongju National University, 56 Gongjudaehak-ro, Gongju-si, Chungcheongnam-do 32588, Korea}
\affil[7]{Chinese Academy of Sciences, National Astronomical Observatories, Key Laboratory of Optical Astronomy}
\affil[8]{National Astronomical Research Institute of Thailand, Chiang Mai 50180, Thailand}

\def\corrauthor{%
Byeong-Cheol Lee, \email{bclee@kasi.re.kr}
}

\def\runningauthor{%
Kim et al.
}

\def\runningtitle{%
A Search for Exoplanets around Northern Circumpolar Stars X
}

\def\keywords{%
stars: individual: \mbox{HD 216595} --- techniques: radial velocities --- stars: planetary systems --- stars: brown dwarf --- stars: AGB and post-AGB
}

\def\abstracttext{%
Detecting planetary companions around evolved stars, particularly asymptotic giant branch (AGB) stars, is challenging due to intrinsic stellar variability such as surface convection, pulsations, and mass loss, which can produce radial velocity (RV) signals that mimic Keplerian motion. We investigate the origin of long-period, low-amplitude RV variations observed in the AGB star HD 216595 based on high-resolution spectroscopic data spanning approximately 16 years obtained with  the Bohyunsan Optical Astronomy Observatory Echelle Spectrograph (BOES) and the Las Cumbres Observatory Network of Robotic Echelle Spectrographs (NRES) instruments. The RV measurements reveal a statistically significant periodic signal at 567 days that can be described by a Keplerian model consistent with a substellar companion. However, no strong correlations are found between the RV variations and stellar activity indicators, including line bisectors, chromospheric activity, and photometric variability, although weak signals at similar timescales are present in some diagnostics. Given the stellar properties of HD 216595 and similarities to previously reported cases, the observed RV variations are likely related to intrinsic stellar processes, although a companion-induced origin cannot be definitively ruled out. Further progress will require improved diagnostics and more sophisticated modeling to disentangle stellar variability from genuine orbital signals.
}

\begin{document}
\jkashead 


\section{Introduction\label{sec:intro}}

Over the past 30 years, high-resolution spectroscopy and stable wavelength references have enabled accurate measurements of stellar radial velocity (RV). Since the 2000s, searching for exoplanets has been extended to evolved stars (\citealt{2002ApJ...576..478F,2003A&A...403.1077K,2003ApJ...597L.157S,2003A&A...398L..19S,2005A&A...437..743H,2007A&A...472..649D,2009ApJ...693.1084W,2011A&A...529A.134L,2020A&A...644A...1T}). 
Giant stars are likely to exhibit more complex RV variations due to various surface processes, such as stellar pulsations, chromospheric activity, spots, and convective cells, which can significantly affect their spectral line profiles. Furthermore, the longer observational baselines typically required for giant stars further complicate the interpretation compared to main-sequence stars.

Proving the existence of low-mass companions around evolved stars, particularly red giant branch (RGB) and asymptotic giant branch (AGB) stars, is extremely challenging because their radii expand by factors of tens to hundreds,
and their luminosities increase by factors of hundreds to thousands compared to the Sun.
Although more than 8,000 exoplanets have been discovered to date,\footnote{The Extrasolar Planets Encyclopaedia: \url{https://exoplanet.eu}
less than 2\% have been found around giant stars.} Furthermore, companions detected around RGB stars are relatively rare, and no confirmed planet has yet been identified around an AGB star; most reported companions in such evolved systems fall within the substellar mass regime (\citealt{2018JKAS...51...17B,2020A&A...638A.148B,2020A&A...644A...1T,2023A&A...678A.106L,2023JKAS...56...35L}).

Also, the study of exoplanets around evolved stars on the RGB and AGB offers a unique opportunity to explore the fate of planetary systems beyond the main-sequence \citep{2003A&A...398L..19S, 2005A&A...437L..31S, 2007A&A...472..657L, 2009ApJ...693..276N, 2023A&A...678A.106L, 2024AJ....167...59X}. 
During these evolutionary phases, strong radiation, stellar winds, and significant mass loss can severely impact the stability of orbiting planets. Nevertheless, detections of exoplanets in such environments provide valuable insights into dynamical evolution \citep{2013MNRAS.432..438A}.

Since 2010, we have been conducting the Search for Exoplanets around Northern Circumpolar Stars (SENS; \citealt{2015A&A...584A..79L}) at the Bohyunsan Astronomical Observatory (BOAO). This survey focuses on an observational study of approximately 400 stars, primarily late-type stars. In this study, we present the HD~216595 observations.
In Section 2, we describe our observations. In Section 3, the stellar properties are presented. Orbital solution and data analysis, and possible origins of RV variations, are presented in Sections 4 and 5, respectively. Finally, Section 6 provides a discussion of our findings.
The appendices include RV measurements obtained in this study (Appendix A) and RV measurements from the literature (Appendix B).

%
   \begin{figure}[t]
   \centering
   \includegraphics[width=8.5cm]{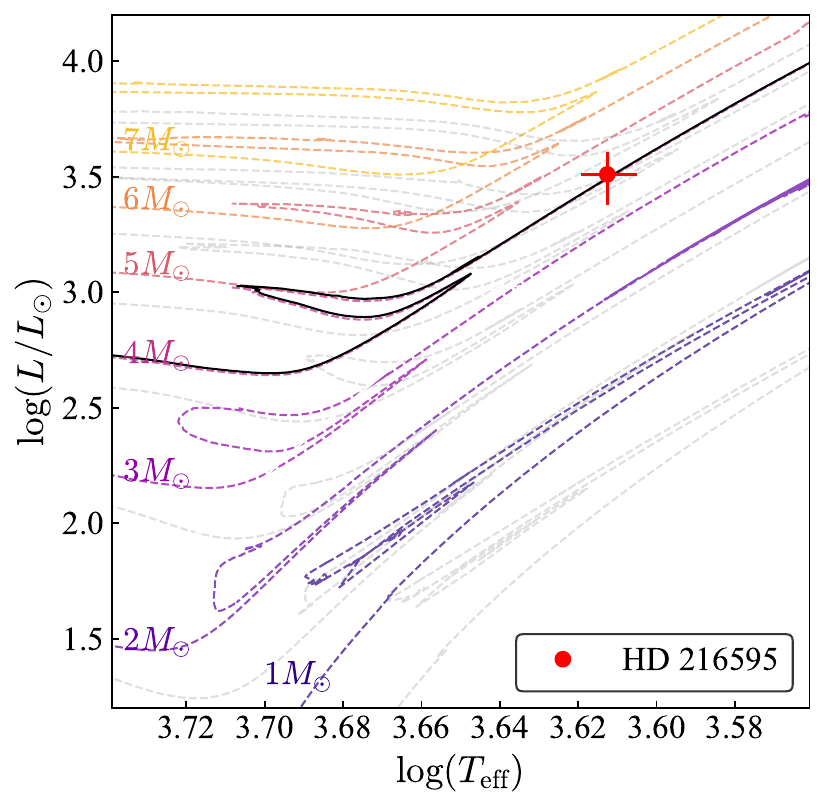}
      \caption{
      Evolutionary stage of HD 216595. The dashed curves represent the MIST evolutionary tracks for different stellar masses. The red cross  symbol indicates the position of HD~216595 for mass of 4.0\,$M_{\odot}$ and metallicity of $-$0.444 $\pm$ 0.267 (black curve).
        }
        \label{f1}
   \end{figure}
%

%
\begin{table*}[t]
\begin{center}
\caption{Stellar parameters.}
\label{tab1}
\begin{tabular}{lccc}
\hline
\hline
	Parameter					& HD 216595   & Reference\\
\hline
Spectral type					& K2        &  \citet{1997yCat.1239....0E}\\
$\textit{$m_{v}$}$ (mag)		& 6.0       &  \citet{1997yCat.1239....0E}\\
$\textit{$M_{V}$}$ (mag)		&  -- 2.5   &  \citet{1997yCat.1239....0E}\\
$\textit{B -- V}$ (mag)			&1.75 $\pm$ 0.02 & \citet{2007AA...474..653V}\\
distance (pc)                   & 504.57$\pm$ 12.17 & \citet{2020yCat.1350....0G}\\
Age (Gyr)						& 0.15$^{+0.12}_{-0.05}$        & MIST\\ 
$\pi$ (mas)					    &  1.98 $\pm$ 0.05   &  \citet{2023AA...674A...1G}\\
$T_{\rm{eff}}$ (K)			   & 3950 $^{+111}_{-153}$ & \citet{2018AA...616A...1G}\\                  
$\rm{[Fe/H]}$				   &  -- 0.444 $\pm$ 0.267  & \citet{2024AA...685A..59M}\\
log $\it g$ (cgs)	           &  0.92 $^{+0.09}_{-0.08}$ &  MIST  \\                	                    
                               & 1.3 $\pm$0.3 &  \citet{2024AA...691A..98K}   \\
$\textit{$R_{\star}$}$ ($R_{\odot}$)  &   112$^{+15}_{-16}$  & MIST\\                            
                                 & 107.06$^{+8.81}_{-5.78}$ &  \citet{2018AA...616A...1G}\\
                                 & 116.58                   & \citet{2023ApJS..268....4F}\\         
$\textit{$M_{\star}$}$ ($M_{\odot}$) &  4.0$^{+0.7}_{-0.9}$  &  MIST \\     
$\textit{$L_{\star}$}$ ($L_{\odot}$) & 3236$^{+812}_{-859}$  & MIST \\  
                                     & 2514.048$\pm$ 115.154  & \citet{2018AA...616A...1G}\\ 
                                     & 4289.63  & \citet{2023ApJS..268....4F}\\               
$P_{\rm{rot}}$ / sin $i$ (days)		 & 2833 $\pm$ 379 & This work\\
\hline
\end{tabular}
\end{center}
\end{table*}

\section{Observations and data reduction} \label{sec:style}
\subsection{BOAO/BOES}
Spectroscopic observations were made from 2010 to 2025 using the fiber-fed high-resolution Bohyunsan Observatory Echelle Spectrograph (BOES; \citealt{2007PASP..119.1052K}) connected to the 1.8 m telescope at BOAO in Korea. The spectra taken with the BOES cover 3,500 ${\AA}$ to 10,500~${\AA}$ over 75 spectral orders. We used a fiber with a diameter of 80 microns that achieves a spectral resolution ($\emph{R}$) of 90,000. In addition, an iodine absorption (I$_{2}$) cell was attached to the front of the BOES to achieve more precise RV measurements. The signal-to-noise ratio (SNR) in the I$_{2}$ region was approximately 100 to 200 with typical exposure times ranging from 600 to 1200~s, depending on atmospheric conditions. The RV standard star $\tau$ Ceti, monitored at BOAO since 2003, has demonstrated a stable RV within 7.5 m s$^{-1}$ \citep{2013A&A...549A...2L}, which is comparable to the typical precision achieved in high-resolution RV surveys.

Standard reduction procedures for raw CCD images were performed using the IRAF software package. Precise RV measurements were made using the I$_{2}$ method and the RVI2CELL package \citep{2007PKAS...22...75H} based on the methods of \citet{1995PASP..107..966V}, \citet{1996PASP..108..500B}, and \citet{2000A&A...362..585E}. Observational summaries are shown in Appendix A (Table~\ref{tab9}).

\subsection{LCO /NRES}
We also performed high-resolution spectroscopy of the star from May 2023 to November 2025 using the Las Cumbres Observatory (LCO; \citealt{2013PASP..125.1031B})  Network of Robotic Echelle Spectrographs (NRES; \citealt{2018SPIE10702E..6CS}) attached to telescopes at McDonald observatory in USA and WISE observatory in Israel. It has a spectral resolution of 53,000 with a wavelength coverage between 3800 ${\AA}$ and 8600 ${\AA}$. The observations were carried out on several nights with exposure times between 500 and 600~s, producing SNRs of approximately 150--200.

The data extracted from the LCO archive were bias and flat-field corrected images processed with the BANZAI pipeline \citep{2018SPIE10707E..0KM}. BANZAI-NRES is designed to handle all data from the NRES of the LCO network.  The pipeline provides the extracted and wavelength-corrected spectra. If the target is a star, it also provides stellar parameters (e.g. effective temperature and surface gravity) and RV measurements. 
The final RVs were measured by combining and averaging the normalized cross-correlation functions (CCF; \citealt{2002A&A...388..632P}) in the 38 orders used to calculate the RV.
And, it shows an average RV errors of $\sim$85 m\,s$^{-1}$ for HD~216595. 
Observational summaries are shown in Appendix A (Table ~\ref{tab10}).

\section{Stellar properties} \label{sec:Stellar properties}
\subsection{Fundamental Parameters}
The basic stellar parameters were obtained from the \textit{HIPPARCOS} catalog \citep{1997yCat.1239....0E} and Gaia database \citep{2018AA...616A...1G}. The star was classified as only K2 spectral type \citep{1997yCat.1239....0E} without luminosity class. Given low effective temperature, surface gravity, large radius, and high luminosity, the star should be giant.
The projected rotational velocity ($v_{\rm rot}\sin i$) was estimated
from line-profile fitting following the method of \citet{2008PASJ...60..781T}.
Using the derived value of $v_{\rm rot}\sin i \simeq 2.0~{\rm km\,s^{-1}}$
and the stellar radius, we estimated a projected rotational period of
 2833 $\pm$  379~{\rm days} for HD~216595.
The basic stellar parameters are summarized in Table~\ref{tab1}.

The RVs of HD~216595 have been measured in several previous studies, where they have been observed to show variations. Their RVs vary within a few hundred m s$^{-1}$. These fluctuations may raise questions about the possibility of planetary or sub-stellar companions. Several previous RV measurements are summarized in Appendix B (Table ~\ref{tab11}).

\subsection{Stellar Characteristics}
We rederived the stellar parameters of the star using the \texttt{isoclassify} package \footnote{\url{https://github.com/danxhuber/isoclassify}}\citep{2017ApJ...844..102H}. First, we applied the Stefan–Boltzmann law to derive the radii of the stars ($R_{\star}$) from posterior probability distributions, using inputs including stellar effective temperature from Gaia DR2 \citep{2018AA...616A...1G}, parallax measurements from Gaia DR3 \citep{2023AA...674A...1G}, the optical photometry ($G_{\rm{BP}}$, and $G_{\rm{RP}}$) from Gaia DR3 \citep{2023AA...674A...1G}, and infrared photometry ($J$, $H$, and $K_{\rm{s}}$) from the Two Micron All-Sky Survey \citep{2006AJ....131.1163S}. Then, we determined various stellar parameters (e.g., $R_{\star}$, $M_{\star}$, $L_{\star}$) from the posterior distributions by integrating over the MESA Isochrones \& Stellar Tracks \citep[MIST,][]{2016ApJ...823..102C} using atmospheric parameters, parallax, and their priors.

The estimated parameters were $R_{\star}$ = 112$^{+15}_{-16}$ $R_{\odot}$, $M_{\star}$ = 4.0$^{+0.7}_{-0.9}$ $M_{\odot}$ and $L_{\star}$ = 3236$^{+812}_{-859}$ $L_{\odot}$ for HD~216595.
The radius is consistent with those of previous studies (\citealt{2018AA...616A...1G,2023ApJS..268....4F}), while the luminosity is within an error margin of approximately 30\% with previous measurements.
The theoretical stellar evolutionary tracks in the Hertzsprung-Russell (H$-$R) diagram of HD~216595 is shown in Fig.~\ref{f1}. HD~216595 is definitively located in the AGB phase on the H-R diagram and evolved from a red clump star.

\begin{figure}
\centering
\includegraphics[width=8.5cm]{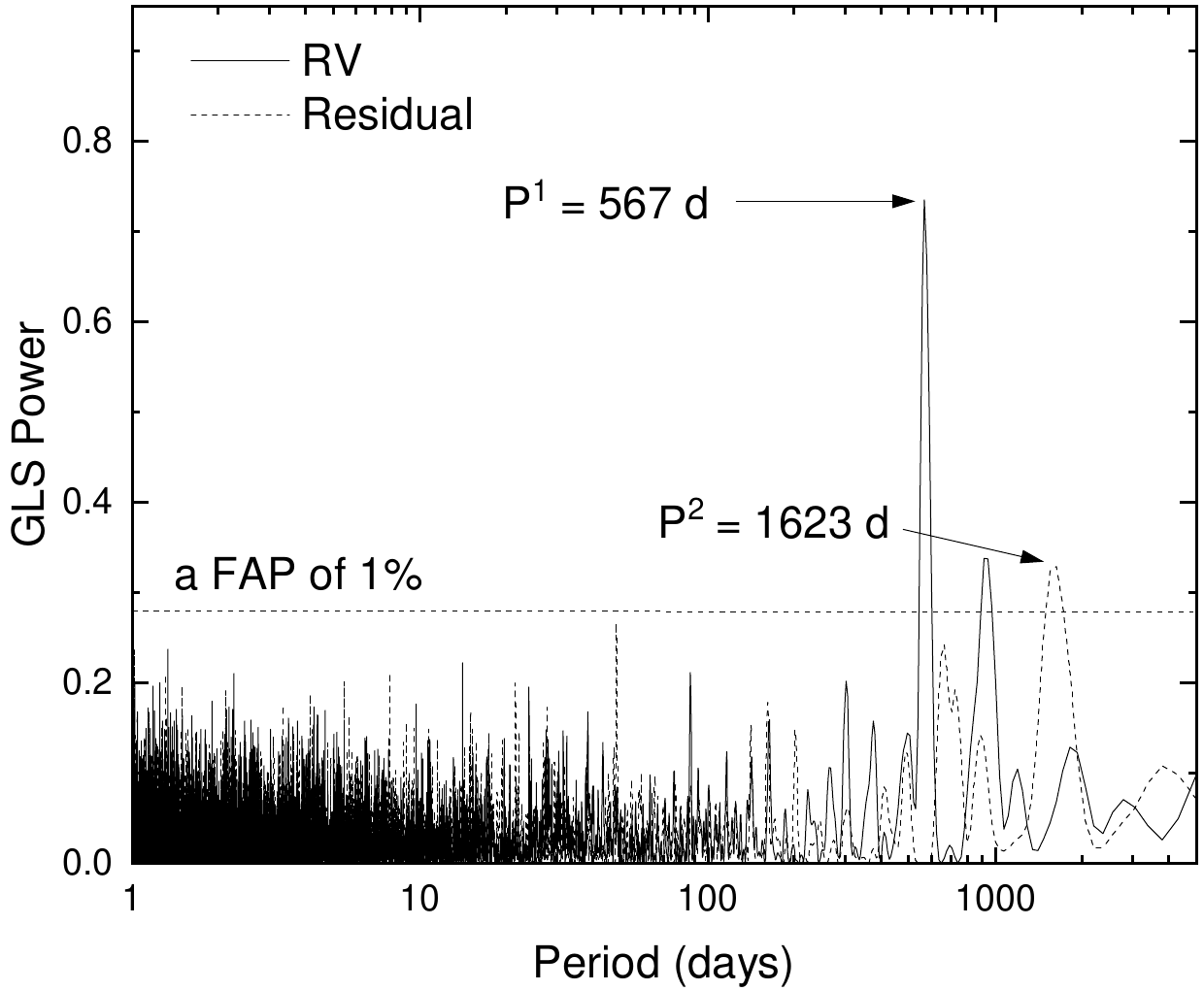}
\caption{The GLS periodograms for RV variations of HD~216595. A strongest peak is 567 days  and
the periodogram for the residual RVs (dashed line) after subtracting the signal with the strongest peak shows a period of 1623 days. The horizontal line  corresponds to a 1\% FAP. \label{f2}}
\end{figure}

%
\section{Orbital solution}
%

\begin{figure}[t]
\centering
\includegraphics[width=8.5cm]{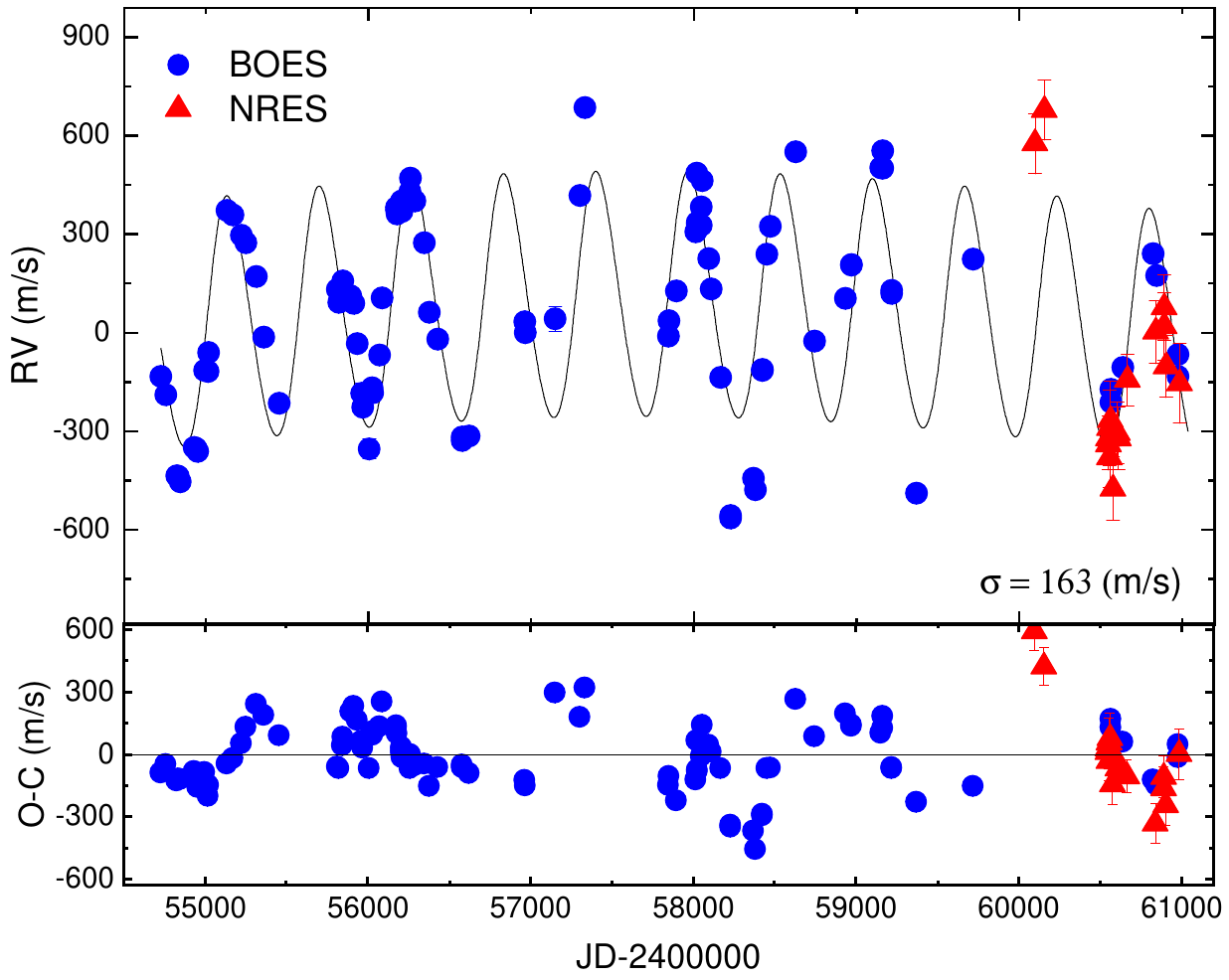}
\caption{RV measurements and residuals for HD~216595.
In top panel, the solid line is the orbital solution with a significant period of 567 days. Bottom panel shows the residual RVs after subtracting the variation corresponding to the significant period.
} \label{f3}
\end{figure}

\begin{figure}[t]
\centering
   \includegraphics[width=8.5cm]{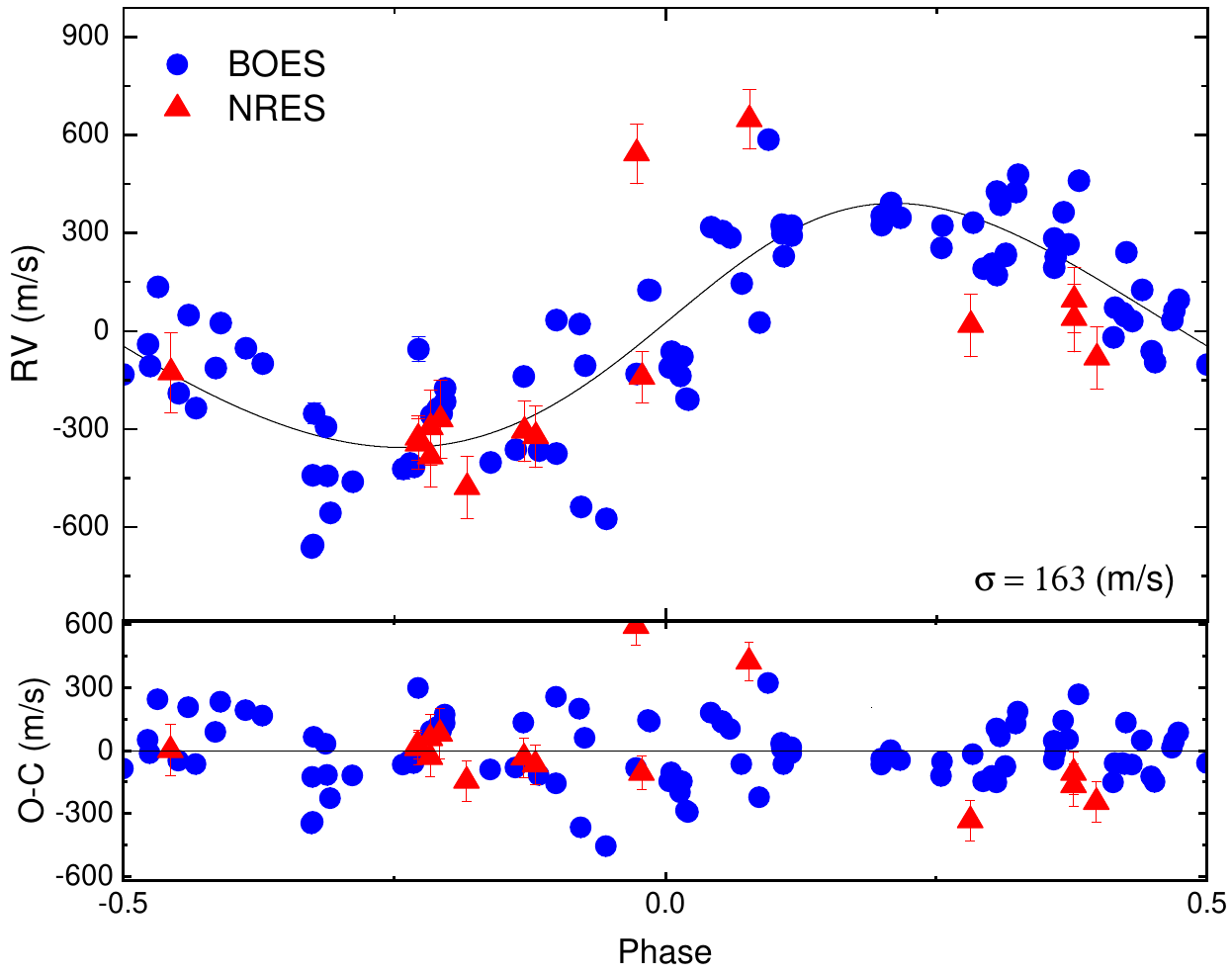}
\caption{Phase diagram with the period of 567 days for HD 216595. \label{f4}}
\end{figure}

   \begin{figure}
   \centering
   \includegraphics[width=8.5cm]{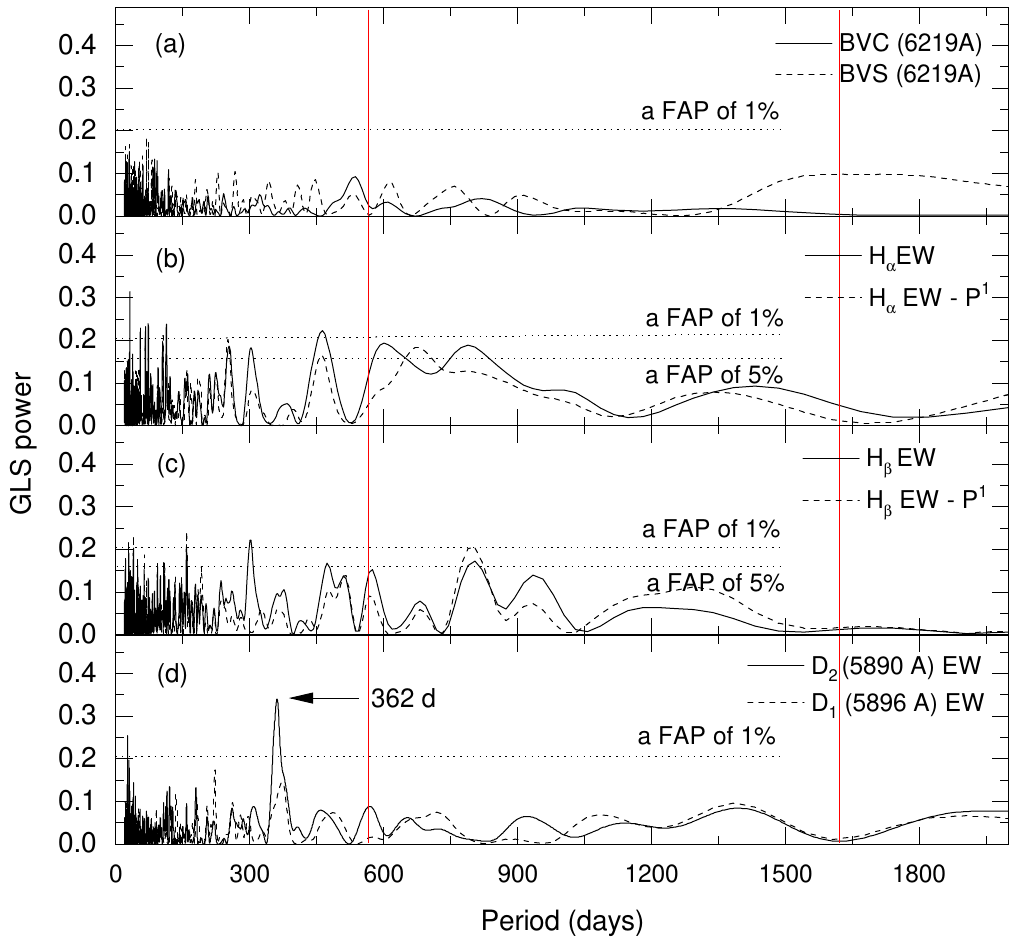}
      \caption{The GLS periodograms of spot indicator and chromospheric activity indicators for HD~216595:
       (a)  periodograms of the BVC and the BVS, (b) periodograms of the H$_{\alpha}$ EW variation, (c)  periodogram of the H$_{\beta}$ EW variation, and (d) periodograms of the Na D$_{1}$ and D$_{2}$ EW variations. The horizontal lines mean corresponds to 1\% and 5\% FAP, respectively.
      The vertical red solid lines  indicate the locations of the periods of 567 and 1623 days, respectively.
        }
        \label{f5}
   \end{figure}
%

We employed the Generalized Lomb-Scargle (GLS; \citealt*{2009A&A...496..577Z}) periodogram to search for periodic signals in the RV data. The GLS yields more accurate frequency estimates and is less sensitive to aliasing effects. Subsequently, orbital parameters were derived using the RVI2CELL package \citep{2007PKAS...22...75H}.

Figure~\ref{f2} presents the GLS  periodogram for HD~216595, revealing a significant periodicity at 567 days. We verified that the detected signal is not associated with the spectral window function or observational cadence. After removing this signal from the  RV data, we applied a pre-whitening procedure \citep{2011A&A...533A...4B} to the residuals revealing an additional signal with a period of 1623 days (Fig.\ref{f2}). The RV time series for HD~216595 is shown in the lower panel of Fig.\ref{f3}, while the phase-folded RV curve for the 567-day period is presented in Fig.~\ref{f4}. The orbital solution corresponding to the 1623-day signal is discussed in Sect. 5.4.

\section{Origin of the RV variations}
Evolved stars frequently exhibit long-period, low-amplitude RV variations, which can originate from multiple astrophysical mechanisms. Identifying the true physical sources of these signals in RGB and AGB stars is particularly challenging because of their extended convective envelopes, intricate internal structures, and the incomplete theoretical framework describing their dynamical behavior. In such stars, intrinsic stellar processes--such as convection, pulsations, and magnetic activity - can induce or obscure RV variations potentially leading to misinterpreted as the signatures of orbiting sub-stellar companions.

\subsection{Rotational Modulation by surface inhomogeneities}  

Surface inhomogeneities (e.g., starspots, chemical abundance anomalies, and magnetically active regions) can induce RV variations through rotational modulation by distorting spectral line profiles, thereby producing quasi-periodic signals that may mimic or obscure planetary signatures.

Rotational modulation associated with surface inhomogeneities produces variable asymmetries in spectral line profiles, which can be quantified using line bisector diagnostics \citep{2001A&A...379..279Q}. To investigate this effect for HD~216595, we computed two standard bisector indicators---the bisector velocity span (BVS) and the bisector velocity curvature (BVC)---from the Fe~I~6219.281\,\AA{} line at two flux levels (40\% and 80\%) of a strong, unblended spectral feature. The generalized Lomb--Scargle (GLS) periodogram of these indicators [Figure~\ref{f5}(a)] shows no significant periodicity, including at the 567-day RV period.

In addition, we analyzed photometric data from the
\textit{Transiting Exoplanet Survey Satellite}
(TESS; \citealt{2015JATIS...1a4003R}) to search for
brightness variations that might be associated with
stellar activity. We used TESS full-frame image (FFI)
light curves obtained from the SPOC/TESS-SPOC pipeline.
After removing data points with non-zero quality flags
and obvious outliers, each observing sector was
normalized independently by its median flux to correct
for instrumental offsets before the light curves were
combined.

The final dataset comprises 98\,807 measurements
spanning approximately 1870\,d
(JD\,2458764.683627--2460636.043436). However,
because the available observations consist of
relatively short TESS sectors that were normalized
independently, the combined light curve is not suitable
for investigating coherent photometric variability on
timescales comparable to the 567-day RV period.
Accordingly, the TESS data were used only to search
for short-timescale photometric variability and do not
provide meaningful constraints on long-period
brightness variations.

Taken together, the absence of significant
periodicities in the bisector indicators provides no
compelling evidence that the observed RV variations
originate from rotational modulation. The available
TESS photometry neither supports nor rules out
low-amplitude photometric variability on timescales
comparable to the RV period. Therefore, while the
current data provide no positive evidence for
rotational modulation, this possibility cannot be
conclusively excluded.

\subsection{Chromospheric Activity Diagnostics}  

Chromospheric activity diagnostics were employed to probe magnetic phenomena in the stellar chromosphere, which is characterized by a positive temperature gradient and enhanced emission features driven by magnetic heating. Different chromospheric indicators trace physical conditions at distinct atmospheric layers: the He I D$_{3}$ line \citep{2021A&A...652A.146L} samples the upper chromosphere and the chromosphere--corona transition region; the H$_{\alpha}$ line \citep{1984ApJ...279..763N,1997A&AS..125..263M} probes the upper chromosphere; the H$_{\beta}$ and Ca II H \& K lines \citep{1984ApJ...279..763N,1997A&AS..125..263M} trace the middle to upper chromosphere; and the Na~D$_{1}$, D$_{2}$ \citep{1989PASP..101..528B,1997A&A...322..266A,2000ApJ...539..858M,2007MNRAS.378.1007D} and Ca II infrared triplet (IRT) lines \citep{2011MNRAS.414.2629M} sample the middle to lower chromosphere.

Among these, the H$_{\alpha}$ and H$_{\beta}$ indices are widely utilized as proxies of chromospheric activity. In particular, the H$_{\alpha}$ line is often preferred for late-type stars, which exhibit increasing flux toward the red spectral region \citep{2007A&A...469..309C}. The relative absence of strong telluric contamination and the narrow hydrogen absorption cores make equivalent width (EW) measurements relatively straightforward. For HD~216595, we measured the EW of the H$_{\alpha}$ line using a passband of $\pm 1.0$ \AA, and the H$_{\beta}$ line using a passband of $\pm 0.8$ \AA, both centered on the line cores to minimize contamination from nearby blends and atmospheric H$_{2}$O absorption.

The periodograms of the H$_{\alpha}$ and H$_{\beta}$ indices show several significant peaks that do not coincide with the dominant RV periodicities. A signal near 567 days is present in both indicators; however, it remains only marginally significant, with a false-alarm probability of $\sim$5\% [Fig.~\ref{f5}(b,c)]. Therefore, while this weak correspondence may suggest a weak correspondence, it does not provide statistically robust evidence for a common origin.

The Ca II H \& K lines, which originate in the chromosphere, often exhibit emission reversals at their line cores as signatures of magnetic activity \citep{1997ApJ...485..319S}. However, the signal-to-noise ratio of our BOES spectra in this wavelength region is insufficient to reliably assess the presence of such features, limiting their diagnostic power in this case.

The Na~D$_{1}$ and D$_{2}$ lines were also analyzed as chromospheric activity indicators.
The Na~D$_1$ (5895.924\,\AA) and Na~D$_2$ (5889.951\,\AA) EWs were measured using 1.0\,\AA-wide passbands centered on each line core.
The resulting EW measurements show no significant periodicity corresponding to the RV signal. Instead, a prominent peak is detected at $\sim$362 days. Given its close proximity to the Earth's annual observing period (365.25 days), this signal is most likely an alias introduced by the seasonal sampling pattern and the observational window function rather than a manifestation of intrinsic stellar variability [Fig.~\ref{f5}(d)].

Taken together, the chromospheric diagnostics does not provide strong or conclusive evidence that the observed RV variations are driven by magnetic activity. Nevertheless, the presence of a weak and marginal signal near the 567-day period suggests that stellar activity cannot be entirely excluded. This ambiguity is consistent with the broader difficulty of distinguishing between intrinsic stellar variability and Keplerian motion in evolved stars, as further discussed in Section~6.

\begin{table}
\begin{center}
\caption{Radial pulsation modes.}
\label{tab6}
\begin{tabular}{lcc}
\hline
\hline
    Parameter                 &                 & Value   \\
\hline
    Fundamental radial pulsation period       & [days]  & 51 $\pm$ 8 \\
    Pulsation RV amplitude   & [m s$^{-1}$]     & 189 $\pm$ 61   \\   
    PLR & [days]      &  $<$ 10   \\ 
    PMR  & [days]     &  $<$ 60 \\      
\hline
\end{tabular}
\end{center}
\end{table}
%

   \begin{figure*}[t]
   \centering
   \includegraphics[width=15cm]{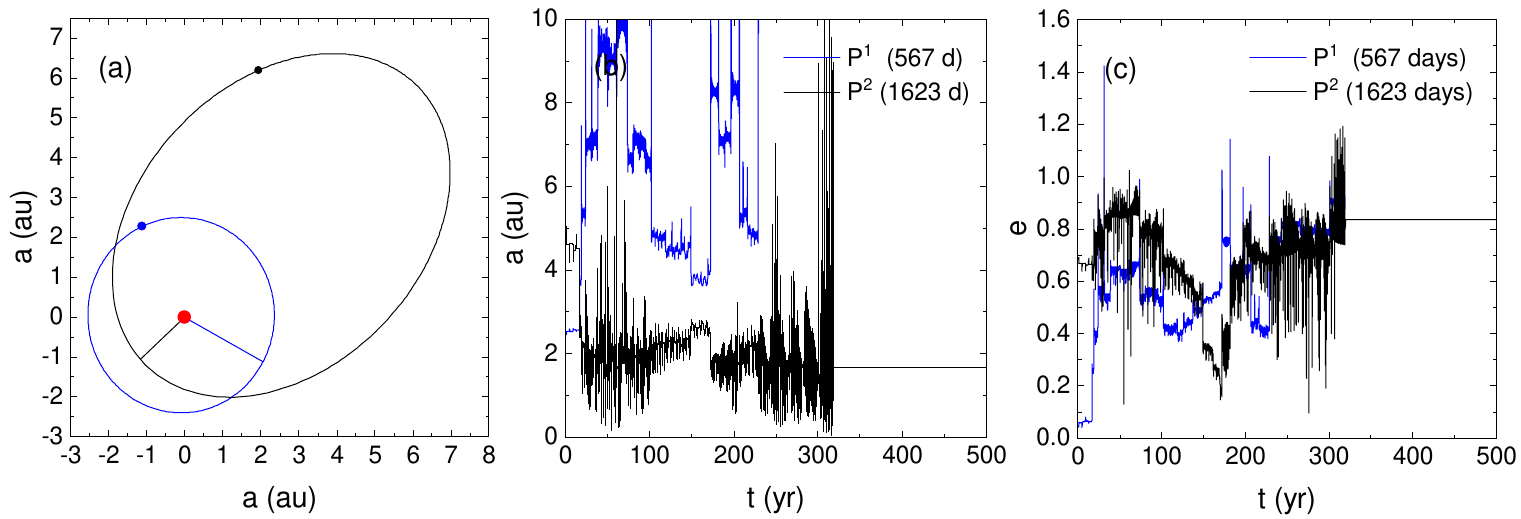}  
      \caption{Assuming the presence of two companions, the orbital evolution of the HD 216595 system for 500 years from the best-fit orbit: (a) orbits of the system, (b) the evolution of the semi-major axes, and  (c) the evolution of the eccentricities.  Blue color indicates the inner companion and black color is the outer companion.}
         \label{f6}
   \end{figure*}
%

\subsection{Stellar pulsation} 
Short-term ($\sim$ days)  RV  variations are primarily caused by radial pulsations. Long-term, low-amplitude RV variations ($\sim$ hundreds of days) may result from non-radial pulsations, rotational modulation, or the presence of low-mass companions. AGB stars commonly exhibit large-amplitude pulsations in fundamental or overtone modes, with periods ranging from tens to hundreds of days. These pulsations drive atmospheric motions that induce RV shifts and can also produce line profile variations, complicating the interpretation of RV signals.
In addition, \citet{2004ApJ...604..800W} reported that approximately 25\%--30\% of pulsating AGB stars exhibit a long secondary period (LSP), typically ranging from 400 to 1500 days. The LSP is considered one of the defining characteristics of AGB stars and is generally about ten times longer than the primary pulsation period, which usually falls in the range of $\sim$30--200 days.

Theoretical Period--Mass--Radius (PMR) relations
\citep{1982ApJ...259..198F} and empirical
Period--Luminosity Relations (PLRs)
\citep{2008MNRAS.386..313W} are commonly used to estimate
pulsation periods in evolved RGB and AGB stars.
The fundamental radial pulsation period and the expected
pulsation-induced RV amplitude were estimated using the
scaling relations of \citet{1995A&A...293...87K}.
For HD~216595, the pulsation periods predicted by the PMR
method range from 18 to 57 days, depending on the adopted
pulsation constant ($Q$), while the PLR method yields
similar timescales differing by only a few days.
These predicted pulsation periods are substantially shorter
than the observed RV period. Therefore, the observed RV
variations are unlikely to be explained by simple radial
pulsation modes, although more complex forms of stellar
variability cannot be entirely excluded.
The pulsation properties derived from the various empirical
and theoretical methods are summarized in
Table~\ref{tab6}.

However, calculating additional pulsation periods for RGB or AGB stars is challenging and remains challenging due to nonlinear pulsations, mass loss, uncertainties in pulsation modes, and the limitations of empirical relations. Simple PMR or PLR models cannot fully account for variations in size and structure. A comprehensive understanding requires advanced multi-wavelength observations and detailed theoretical modeling.

In our study, we identify a strong period of $\sim$1870 days in the photometric data of HD~216595, consistent with the LSP regime described by \citet{2004ApJ...604..800W}.  However, although this period generally falls into the category of LSP photometric cycles observed in AGB stars, no clear evidence is found linking them to chromospheric activity. 
Moreover, the long-term RV variations of our targets differ significantly from the typical magnitude variations (several magnitudes) or RV amplitude variations (a few km s$^{-1}$) commonly observed in LSPs \citep{2009MNRAS.399.2063N}, indicating that the observed RV variations are not fully consistent with typical LSP behavior. Ultimately, the nature of the secondary long-term RV fluctuations in our target remains unresolved.

\subsection{Low-mass companion} 
The observed periodic RV variations exhibit a coherent Keplerian signal, which can be modeled by a low-mass companion scenario. According to the best-fit orbital solution, this object can be interpreted as a candidate sub-stellar companion with a minimum mass of $39.3^{+5.2}_{-5.0}\ M_{\mathrm{J}}$, an orbital period of $567.29^{+0.05}_{-0.05}$ days, and a semi-major axis of $2.14^{+0.14}_{-0.14}$ AU.

\citet{2019AA...623A..72K} analyzed astrometric data from both \textit{Gaia} DR2 and \textit{HIPPARCOS} to constrain the long-term motion of stars appearing in both catalogs. Assuming a random distribution of orbital inclinations, they derived a mean inclination of $60^{+21}_{-27}$ degrees. Based on this, they inferred a secondary companion around HD~216595 (adopting a stellar mass of $6.02~M_\odot$), with an estimated mass of $68.20^{+44.36}_{-40.72}~M_{\mathrm{J}}$ at a projected separation normalized to 1~AU.

Although the companion-mass estimates derived from the RV and astrometric approaches are not identical, the discrepancy can be reasonably attributed to differences in the adopted stellar mass, orbital geometry, and inclination assumptions. Within these uncertainties, the inferred properties are broadly consistent, and thus the presence of a low-mass companion associated with the 567-day signal cannot be ruled out.

In addition to this primary signal, a second RV variation with a period of $\sim1623$ days was also explored as a possible companion candidate. Under a two-companion interpretation, the first candidate would have a mass of approximately $39~M_{\mathrm{J}}$ with an orbital period of 567 days at a separation of 2.1 AU, while the second candidate would have a mass of about $30~M_{\mathrm{J}}$ with an orbital period of 1623 days at a separation of 4.4 AU.

However, for multi-body systems, long-term dynamical stability is a necessary condition for the viability of any proposed orbital configuration. To test this, we performed numerical orbital integrations using a customized version of the Simba Simple N-Body Algorithm \citep{1998AJ....116.2067D} implemented in the \texttt{Exo-Striker} toolbox \citep{2019ascl.soft06004T}. A total of 5,000 parameter sets randomly drawn from the posterior distributions were integrated for 500 years with a time step of 0.2 days. As shown in Figure~\ref{f6}, the two-companion configuration is highly unstable, with strong mutual interactions leading to orbital disruption within $\sim300$ years.

Therefore, the 1623-day RV variation is unlikely to originate from a genuine Keplerian companion. This result weakens the interpretation of the \emph{secondary} signal as having a Keplerian origin. However, it does not preclude a companion-induced interpretation for the dominant 567-day RV signal, which remains consistent with a Keplerian model but cannot be unambiguously distinguished from intrinsic stellar variability.
The derived orbital parameters are summarized in Table 3.

\begin{table}[!]
\renewcommand{\thetable}{\arabic{table}}
\centering
\caption{Orbital solution for a significant period by the MCMC simulation for HD 216595.} \label{tab3}
\begin{tabular}{lcc}
\hline
\hline
Parameter	& HD 216595 b   \\
\hline
P (days)			& 567.29$^{+0.05}_{-0.05}$	\\
K (m s$^{-1}$)		& 381.06 $^{ +0.09}_{-0.08}$	 \\
$e$					& 0.09  $^{ +0.01}_{-0.01}$  \\
$T_{periastron}$ (JD) & 2454484.9937 $\pm$ 52.8897\\
$\omega$ (deg)		& 303.63 $^{+0.02}_{-0.05}$ \\
$m$ sin $i$ ($M_{J}$)& 39.3$^{+5.2}_{-5.0}$   \\
$a$ (AU)         	 & 2.14$^{+0.14}_{-0.14}$       \\
$Slope$ Linear trend (m s$^{-1} d^{-1}$) & 0.0564 $\pm$ 0.0035   \\
RV jitter amplitude (m s$^{-1}$)   & 167.3 $\pm$ 10.0\\
$N_{obs}$                & 114  \\
$chi^2$ (m s$^{-1}$) &	97.4	 \\
rms (m s$^{-1}$) &	160.0	 \\
\hline
\end{tabular}
\end{table}


\subsection{Surface convection, granulation, and stellar winds}

Evolved stars, particularly those on the AGB, are expected to exhibit large-scale convective cells and granulation patterns that introduce significant surface inhomogeneities. These effects distort spectral line profiles and can induce apparent RV shifts unrelated to orbital motion. Consequently, the observed RV variability may arise from a combination of gravitationally bound companions and intrinsic stellar processes, with the latter contributing to stellar jitter that can mimic or obscure Keplerian signals.

State-of-the-art three-dimensional (3D) radiative hydrodynamic simulations have demonstrated that convection-driven variability can reproduce observed surface structures and is broadly consistent with interferometric constraints \citep{2010A&A...515A..12C, 2011ApJ...741..119M, 2020ApJ...898...24R, 2020A&A...635A.146S}. These simulations further indicate that convection-induced RV signals can exhibit amplitudes and temporal coherence comparable to those expected from orbiting companions. These processes are therefore expected to produce RV signals with amplitudes and timescales comparable to the observed variations in HD~216595, further complicating the interpretation of the Keplerian signal and reinforcing the intrinsic degeneracy between stellar variability and orbital motion.

In addition, AGB stars experience substantial mass loss through stellar winds, which can further perturb the outer atmosphere and introduce additional RV variability via spectral asymmetries or circumstellar absorption features \citep{2003A&A...409..715W, 2018A&A...615A..28D}. These effects may contribute to long-period or quasi-periodic signals, complicating their interpretation in terms of companions.

In the context of our analysis, these considerations imply that caution is required when attributing RV signals to Keplerian motion, particularly for long-period or low-amplitude variations. While companion-induced interpretations remain viable for some signals, alternative explanations related to convection, granulation, or stellar winds cannot be excluded. A more comprehensive treatment incorporating these processes will be necessary to fully disentangle their contributions, and will be explored in future work.

\section{Discussion\label{sec:dis}}
Despite the statistical significance of the Keplerian fit to the RV data of HD~216595, its physical interpretation remains uncertain in the context of AGB stars. The star’s large radius ($R > 20~R_\odot$) and high luminosity ($\log(L/L_\odot) > 2.5$) place it in a regime where spurious RV signals with periods of $\sim$500--600 days are commonly observed \citep{2018A&A...619A...2D}. These signals are generally attributed to intrinsic stellar processes, such as non-radial pulsations and large-scale convective motions \citep{2015MNRAS.452.3863S, 2021ApJS..256...10D, 2024A&A...689A..91S}. Evolved stars in this parameter space are known to exhibit RV variations with amplitudes of $\sim$100--500~m~s$^{-1}$ and periods of several hundred days without requiring planetary companions. Such intrinsic variability complicates the interpretation of RV signals and fundamentally limits the detection of planetary companions around AGB stars.

The system considered here lies within this regime. Its radius ($R \gtrsim 100~R_\odot$), luminosity ($\log(L/L_\odot) \sim 3.5$), and RV period (567 days) are compatible with those of AGB stars exhibiting intrinsic RV variability. The observed RV semi-amplitude ($K \sim 150$--200~m~s$^{-1}$; see Fig.~\ref{f3}) is also consistent with the range expected from convection- or pulsation-driven variability \citep{2024A&A...689A..91S}.

Based on these considerations, together with our analysis of activity diagnostics, three interpretations remain viable: (i) intrinsic variability driven by pulsations and convection, (ii) a companion-induced Keplerian signal, and (iii) chromospheric activity contributing to the observed variability. However, given the limitations of RV diagnostics in AGB stars, the current data do not allow a robust discrimination among these scenarios.

\begin{itemize}
    \item \textbf{Intrinsic stellar variability:}  
    Pulsations, long secondary periods (LSPs), and large-scale convection are expected to induce quasi-periodic RV variations with amplitudes of up to several hundred m~s$^{-1}$ on timescales comparable to the observed period \citep{2015MNRAS.452.3863S, 2024A&A...689A..91S}.
    
    \item \textbf{Companion-induced origin:}  
    The RV signal can be described by a coherent Keplerian model corresponding to a low-mass companion ($\sim 39~M_{\rm J}$, $\sim 2.1$~AU; Sect.~5.4). However, intrinsic stellar variability may mimic such signals in AGB stars.
    
    \item \textbf{Chromospheric activity:}  
    Weak variability is detected in several activity indicators near the proposed period [Fig.~\ref{f5}(b,c)], although not at a statistically significant level (FAP $\sim$ 5\%).
\end{itemize}

In summary, the observed RV signal lies in a regime where all three mechanisms are plausible, and the available diagnostics do not provide conclusive evidence in favor of any single interpretation. Further progress will require improved modelling of stellar variability and coordinated, high-precision observations to disentangle intrinsic stellar signals from genuine orbital motion.


\acknowledgments
BCL acknowledges partial support by the KASI (Korea Astronomy and Space Science Institute) grant
2026-1-904-01 and acknowledge support by the National Research Foundation of Korea (NRF) grant funded by the Korea government (MSIT) (Grant No. RS-2021-NR058435).
SHG acknowledges support by the National Natural Science Foundation of China (NSFC) (grants Nos. 10373023, 10773027 and U1531121), support by the Yunnan Fundamental Research Project (grant No. 202305AS350009), and the science research grant from the China Manned Space Project.
BL acknowledges support by the NRF grant funded by MSIT (Grant No. RS-2022-NR072247).
HYT appreciates the support by the EACOA/EAO Fellowship Program under the umbrella of the East Asia Core Observatories Association.
MGP was supported by the Basic Science Research Program through the NRF funded by the Ministry of Education (RS-2019-NR045193, RS-2018-NR031074) and KASI under the R\&D program supervised by the Ministry of Science, ICT and Future Planning.
HYC was supported by Basic Science Research Program through the National Research Foundation of Korea (NRF) funded by the Ministry of Education (RS-2018-NR031074).


 \bibliography{jkas-ref}

%


\begin{appendix} 
\section{RV measurements}
In this appendix, we present all observational data collected with the BOES and the NRES. We list the observation dates Julian (JD), the radial velocities (RV), and uncertainly ($\pm \sigma$), respectively.

\begin{table*}
\renewcommand{\thetable}{\arabic{table}}
\centering
\caption{Relative RV measurements for HD 216595 from January 2010 to November 2025 using the BOES.} \label{tab9}
\begin{tabular}{ccccccccc}
\hline
\hline
JD & RV  & $\pm \sigma$ & JD & RV  & $\pm \sigma$  & JD & RV  & $\pm \sigma$\\
$-$2,400,000 &{m\,s$^{-1}$}& {m\,s$^{-1}$}& $-$2,400,000 & {m\,s$^{-1}$} & {m\,s$^{-1}$} & $-$2,400,000 & {m\,s$^{-1}$} & {m\,s$^{-1}$}\\
\hline
                                        
54726.239981  &   -202.8   &    12.8 &   56173.129277  &    312.0  &     15.6 &   58093.137748  &    156.7   &    17.4  \\
54755.184179  &   -258.2   &    16.1 &   56177.142862  &    292.8  &     10.1 &   58109.007343  &     63.6   &    14.6  \\
54825.030733  &   -505.3   &    16.4 &   56203.940706  &    332.9  &     16.7 &   58165.910570  &   -206.1   &    24.5  \\
54832.944709  &   -506.6   &    12.1 &   56204.146553  &    327.0  &     17.1 &   58226.324832  &   -634.4   &    19.3  \\
54845.957997  &   -523.2   &    14.4 &   56204.194204  &    305.2  &     20.8 &   58227.317960  &   -627.2   &    16.5  \\
54931.312270  &   -419.2   &    13.4 &   56204.209968  &    317.0  &     13.7 &   58367.203050  &   -513.1   &    15.2  \\
54943.296515  &   -421.9   &    18.6 &   56209.140661  &    329.8  &     17.5 &   58380.354829  &   -548.3   &    26.6  \\
54952.314336  &   -430.7   &    16.2 &   56209.150047  &    300.3  &     15.1 &   58422.114459  &   -181.7   &    16.5  \\
54994.196612  &   -183.7   &    20.7 &   56256.131665  &    333.3  &     14.5 &   58423.168103  &   -184.4   &    17.6  \\
55017.265345  &   -187.7   &    13.5 &   56256.139107  &    361.1  &     14.9 &   58451.116550  &    170.5   &    18.1  \\
55018.220335  &   -129.3   &    15.2 &   56261.084687  &    401.6  &     18.6 &   58472.921106  &    254.3   &    16.7  \\
55131.934559  &    303.4   &    17.8 &   56288.000040  &    332.9  &     13.4 &   58627.192605  &    482.7   &    12.4  \\
55169.966177  &    290.3   &    14.2 &   56346.390057  &    205.1  &     19.9 &   58743.229045  &    -95.0   &    14.2  \\
55219.909165  &    227.7   &    14.5 &   56377.357643  &     -6.9  &     12.1 &   58933.332648  &     35.5   &    22.8  \\
55249.904437  &    204.9   &    11.6 &   56426.311527  &    -89.5  &     12.9 &   58969.265446  &    138.8   &    17.9  \\
55311.320199  &    102.0   &    14.5 &   56576.936908  &   -387.3  &     18.6 &   58970.263133  &    137.2   &    17.2  \\
55357.304873  &    -83.2   &    14.7 &   56578.949508  &   -398.2  &     15.2 &   59151.112632  &    434.1   &    16.3  \\
55454.144149  &   -284.8   &    14.6 &   56618.980314  &   -383.6  &     13.6 &   59161.138449  &    431.8   &    16.4  \\
55811.067046  &     62.3   &    11.7 &   56963.996813  &    -35.9  &     16.3 &   59162.208983  &    485.6   &    17.2  \\
55820.149133  &     22.7   &    12.2 &   56965.980508  &    -68.8  &     11.6 &   59217.007553  &     59.1   &    20.9  \\
55842.067750  &     55.2   &    14.7 &   57148.288954  &    -27.4  &     37.7 &   59218.007501  &     50.2   &    17.7  \\
55844.269040  &     88.2   &    17.0 &   57301.039072  &    349.3  &     14.4 &   59370.253225  &   -558.7   &    13.9  \\
55894.174940  &     43.1   &    18.2 &   57331.075632  &    617.5  &     11.9 &   59718.171039  &    154.7   &    18.6  \\
55911.055526  &     19.5   &    18.1 &   57846.339967  &    -80.2  &     15.5 &   60564.049310  &   -241.4   &    12.6  \\
55933.000252  &   -102.9   &    17.1 &   57847.341855  &    -32.3  &     23.1 &   60564.090978  &   -281.1   &    13.5  \\
55959.945862  &   -254.3   &    31.8 &   57893.271100  &     58.3  &     14.4 &   60636.965925  &   -175.8   &    26.0  \\
55965.931483  &   -296.4   &    20.2 &   58014.936646  &    237.8  &     12.2 &   60823.212995  &    171.9   &    23.2  \\
56006.362244  &   -423.4   &    29.6 &   58018.967605  &    416.8  &     15.8 &   60845.211451  &    105.4   &    23.1  \\
56024.342082  &   -237.6   &    14.8 &   58021.956031  &    261.5  &     16.4 &   60975.884886  &   -135.9   &    20.2  \\
56026.326179  &   -253.1   &    15.3 &   58021.963416  &    267.7  &     15.7 &   60976.966472  &   -202.8   &    23.1  \\
56069.274339  &   -136.7   &    10.3 &   58047.201521  &    314.0  &     17.2 &   60993.967902  &   -284.7   &    16.9  \\
56086.211623  &     36.7   &    10.9 &   58047.957363  &    257.4  &     20.1 &   61007.925809  &   -202.3   &    21.0  \\
56173.012954  &    304.4   &    14.6 &   58051.993484  &    394.8  &     17.7 &                 &            &          \\

\hline
\end{tabular}
\end{table*}

\begin{table*}
\renewcommand{\thetable}{\arabic{table}}
\centering
\caption{RV measurements for HD 216595 from May 2023 to November 2025 using the NRES.} \label{tab10}
\begin{tabular}{lccc}
\hline
\hline
BJD & RV  & $\pm \sigma$ & site\\
$-$2,400,000 &{km\,s$^{-1}$}& {km\,s$^{-1}$} & \\
\hline
60096.91104&    -6.913&	0.092&	McDonald \\
60155.71573&	-6.811&	0.091&	McDonald \\
60549.69550&	-7.830& 0.078&	McDonald\\
60549.70317&	-7.812&	0.069&	McDonald \\
60556.36983&	-7.870&	0.093&	Wise \\
60556.37749&	-7.780&	0.115&	Wise \\
60561.43134&	-7.756&	0.120&	Wise \\
60575.31531&	-7.965&	0.096&	Wise \\
60605.36160&	-7.795& 0.093&	Wise \\
60611.31522&	-7.812&	0.095&	Wise \\
60666.65026&	-7.634&	0.078&	McDonald \\
60838.43504&	-7.487& 0.096&	Wise\\
60892.31046&	-7.413&	0.100&	Wise\\
60892.37984&	-7.470&	0.103&  Wise\\
60904.26048&	-7.592& 0.095&	Wise\\
60987.65357&	-7.644& 0.122& McDonald \\
\hline
\end{tabular}
\end{table*}

\section{RV measurements in literature}
In this appendix, we present the RV measurements of HD~216595 recorded in literature.

%
\begin{table*}
\begin{center}
\caption{RV measurements of \mbox{HD 216595} recorded in literature.}
\label{tab11}
\begin{tabular}{lr}
\hline
\hline
    RV  (km s$^{-1}$)                  &  References      \\
\hline
    -- 7.3            & \citet{1933PDAO....6..149H}    \\ 
    --7.0 $\pm$ 5.0   & \citet{1953GCRV..C......0W}    \\     
    -- 7.0            & \citet{1980BICDS..19...74O}    \\ 
    -- 7.9           &  \citet{1996yCat.3192....0F}  \\      
    -- 7.0          & \citet{1999VeARI..35....1W} \\
    -- 7.41 $\pm$ 0.20  & \citet{2005AA...430..165F}    \\
    -- 7.5 $\pm$ 0.8  & \citet{2006ARep...50..733B}    \\     
    -- 7.4 $\pm$ 0.3  & \citet{2006AstL...32..759G}   \\      
    -- 7.44 $\pm$ 0.24  & \citet{2007AN....328..889K}    \\ 
    -- 7.9            &  \citet{2007AA...463..783V}  \\   
    -- 7.04 $\pm$ 0.20  & \citet{2018AA...616A...1G}    \\  
    -- 6.436 $\pm$ 0.20  &  \citet{2019AA...623A..72K}  \\       
    -- 7.039 $\pm$ 0.204  & \citet{2021ApJS..254...42B}    \\     
    -- 7.71 $\pm$ 0.15  & \citet{2018AA...616A...1G}    \\ 
    -- 7.058 $\pm$ 0.387  & \citet{2022AA...659A..95T}    \\     
    -- 7.04            & \citet{2023AA...669A..15Y}    \\
\hline
\end{tabular}
\end{center}
\end{table*}
%
\end{appendix}

\end{document}